\documentclass[twocolumn,           
               showpacs,            
               preprintnumbers,     
               aps,                 
               prd,          	    
               a4paper,             
               superscriptaddress,  
               nofootinbib,         
               tightenlines,        
               floats,floatfix      
               ]{revtex4}

\usepackage{graphicx}
\usepackage{latexsym}
\usepackage{amsmath,amssymb}        
\usepackage[draft=false]{hyperref}

\begin{document}



\title{Curvaton reheating: an application to braneworld inflation}
\author{Andrew R.~Liddle}
\affiliation{Astronomy Centre, University of Sussex, Brighton BN1 9QJ, United 
Kingdom}
\author{L. Arturo Ure\~{n}a-L\'{o}pez}
\affiliation{Astronomy Centre, University of Sussex, Brighton BN1 9QJ, United 
Kingdom}
\affiliation{Instituto de F\'{\i}sica de la Universidad de Guanajuato, A.
P. E-143, C.P. 37150, Le\'{o}n, Guanajuato, M\'{e}xico}
\date{\today}
\pacs{ 98.80.Cq, 98.70.Vc \hfill astro-ph/0302054}
\preprint{astro-ph/0302054}


\begin{abstract}
The curvaton was introduced recently as a distinct inflationary mechanism for 
generating adiabatic density perturbations. Implicit in that scenario is that 
the curvaton offers a new mechanism for reheating after inflation, as it is a 
form of energy density not diluted by the inflationary expansion. We consider 
curvaton reheating in the context of a braneworld inflation model, {\em steep 
inflation}, which features a novel use of the braneworld to give a new mechanism 
for ending inflation. The original steep inflation model featured reheating by 
gravitational particle production, but the inefficiency of that process brings 
observational difficulties. We demonstrate here that the phenomenology of steep 
inflation is much improved by curvaton reheating. 
\end{abstract}

\maketitle


\section{Introduction}
\label{one}

The curvaton \cite{lyth1,lyth2} is an interesting new proposal for explaining 
the observed large-scale adiabatic density perturbations in the context of 
inflation. Perturbations are acquired by a field other than the inflaton during 
inflation, whose energy density is subdominant during inflation but which comes 
to dominate sometime afterwards, at which point its initial isocurvature 
perturbations convert to adiabatic \cite{moll,lyth1} (see also \cite{TT}). Decay 
of the curvaton to 
conventional matter locks the perturbations in as adiabatic, though there can be 
complex phenomenology if some species have already decoupled from the thermal 
bath by the time this happens \cite{lyth2,ES}.

An aspect of this process which has been implicit in most papers is that this 
process also offers a new mechanism of reheating. The first 
detailed discussion of this aspect appeared recently in Ref.~\cite{FL} by Feng 
and Li, who named it {\em curvaton reheating}. The curvaton is a type of matter 
whose energy 
density is not diluted during inflation, and some or all of the present material 
in the Universe may have survived inflation via the curvaton. This reheating 
mechanism complements the other two known mechanisms, the standard one being 
decay of the inflaton energy density into conventional matter and the 
rarely-used alternative being gravitational particle production at the end of 
inflation \cite{gravpart}.

In this paper we will explore an application of curvaton reheating to a 
particular model of braneworld inflation known as {\em steep inflation} 
\cite{cope}. This models brings the inflationary epoch to an end in a novel way. 
The potential chosen (for example a steep exponential) is so steep that it would 
not drive inflation in the standard cosmology, but it may do so in the presence 
of high-energy braneworld corrections to the Friedmann equation, for example as 
found in the Randall--Sundrum Type II scenario. As the energy density decreases, 
those corrections become unimportant, and the scalar field makes a transition 
into a kinetic energy dominated regime bringing inflation to an end. As the 
inflaton survives this process without decay, an alternative reheating mechanism 
is required.

In Ref.~\cite{cope}, it was proposed that reheating took place via gravitational 
particle production \cite{gravpart,MJ1}. However this brings problems to the 
scenario. Because of 
the inefficiency of the process, there is a long kinetic energy dominated 
regime, which allows short-wavelength gravitational waves to reach excessive 
amplitudes \cite{sahni}. There are likely also to be problems with the low 
reheat temperature achieved, which is barely sufficient to allow standard 
nucleosynthesis to proceed. The scenario also suffers from a separate problem, 
which is that it tends to predict perturbation spectra which are too far from 
scale-invariance, and too high an amplitude of long-wavelength gravitational 
waves; for the exponential potential the predicted spectra are in conflict with 
the most recent observations.

In this paper, we propose that implementation of curvaton reheating in place of 
gravitational particle production can redress those difficulties. Curvaton 
reheating can be much more efficient, allowing a high reheat temperature and 
preventing short-wavelength gravitational wave dominance. Further, adiabatic 
perturbations produced via the curvaton may be closer to scale-invariance, and 
in particular the contribution of gravitational waves to the microwave 
anisotropies may be considerably suppressed. Two related papers appeared as we 
were completing this work; Ref.~\cite{FL} discuss curvaton reheating in a 
quintessential inflation model \cite{PV} in the standard cosmology, while 
Ref.~\cite{Dim} 
considers adding the curvaton to the steep inflation model, but focussing on 
particle physics model building aspects rather than observational phenomenology 
and constraints.

A summary of this paper is as follows. In Section~\ref{two}, we review the main 
features of steep inflation in the braneworld scenario. We present only the main 
results of these kind of models, including the production of primordial 
gravitational 
waves. In Section~\ref{three}, we study the evolution of a curvaton field and 
its perturbations in steep inflation, and we focus our attention on all the 
constraints to be applied upon the free parameters of the model. We consider 
separately the cases that arise depending on whether the curvaton field 
dominates after or 
before it decays into radiation. In Section~\ref{four} we put all constraints 
together and analyze the parameter space of the model of curvaton field $+$ 
steep inflation which satisfies all the requirements. In the final section we 
draw some conclusions and topics to be investigated in future publications.

\section{The Steep Inflation model}
\label{two}

In this section we give a brief description of an inflationary model in the 
braneworld scenario, assuming that the inflaton field is a minimally coupled 
real scalar field $\phi$ endowed with an exponential potential of the form
\begin{equation}
V(\phi)=V_0 e^{-\alpha \sqrt{\kappa_0}\phi} \, , \label{2e1}
\end{equation}
where $V_0, \alpha$ are free parameters and $\kappa_0=8\pi G = 8 \pi  
m^{-2}_{{\rm Pl}}$ with $m_{\rm Pl}$ the Planck mass. More details of the model 
can be found in Refs.~\cite{cope,sahni}. A somewhat related idea has been 
discussed in Ref.~\cite{guendel}.

\subsection{Inflationary dynamics}

In the $(4+1)$-dimensional brane scenario inspired by the Randall--Sundrum 
model, 
the standard Friedman equation is modified to
\begin{equation}
H^2 = \frac{\kappa_0}{3} \rho \left( 1+\frac{\rho}{2\lambda_{{\rm b}}} \right) 
\, ,
\label{2e2}
\end{equation}
where $\lambda_{{\rm b}}$ is the brane tension and, taking the 
inflaton field 
to be the dominant component, $\rho = \dot{\phi}^2/2+V(\phi)$. The scalar 
field is confined to the brane so that its equation of motion is the usual 
Klein--Gordon one
\begin{equation}
\ddot{\phi} + 3H \dot{\phi} + \frac{dV}{d\phi} = 0 \, , \label{2e3}
\end{equation}
where dots denote temporal derivatives.

For large energies $\rho \gg \lambda_{{\rm b}}$, the inflaton field experiences 
an 
increased friction as it rolls down its potential, and then one can expect 
inflation to occur even for steep potentials (i.e. even for $\alpha \gg 1$). 
This can be seen from the slow-roll parameters for the exponential potential 
\cite{MWBH}, 
which for large energies read
\begin{equation}
\epsilon = \eta = 2\alpha^2 \, \frac{\lambda_{{\rm b}}}{V} \, . \label{2e4}
\end{equation}
and thus, inflation can occur whenever $V > 2\alpha^2 \lambda_{{\rm b}}$.

The equations of motion can be solved exactly in the slow-roll approximation, 
from which we find the following expressions for the scale factor $a$, the 
Hubble factor $H$ and the inflaton field $\phi$, respectively,
\begin{eqnarray}
\ln  \frac{a}{a_{{\rm i}}}  &=& ({\cal N}+1)-e^{-H_{{\rm 
f}}(t-t_{{\rm f}})} \, , 
\label{2e5} \\
H &=& H_{{\rm f}} e^{-H_{{\rm f}}(t-t_{{\rm f}})} \, , \label{2e6} \\
\alpha \sqrt{\kappa_0} \phi(t) &=& H_{{\rm f}} (t-t_{{\rm f}}) - \ln  
\frac{V_{{\rm f}}}{V_0}  \, . \label{2e7}
\end{eqnarray}
Here `i(f)' indicates the initial (final) values of the different quantities 
during inflation, and ${\cal N}$ is the number of $e$-foldings as can be seen 
from Eq.~(\ref{2e5}). Observe that, in the braneworld scenario, the slow-roll 
approximation for the exponential potential is a much better approximation than 
in standard cosmology. 

Inflation ends by violation of the slow-roll conditions, when $\epsilon \simeq 
\eta \sim 1$, and so the value of the potential energy at the end of inflation 
is $V_{{\rm f}} \simeq 2 \alpha^2 \lambda_{{\rm b}}$. Using 
Eqs.~(\ref{2e5})--(\ref{2e7}), we find that the initial and final values of the 
Hubble factor and the potential energy, respectively, are related through 
\begin{equation}
\frac{H_{{\rm i}}}{H_{{\rm f}}} = \frac{V_{{\rm i}}}{V_{{\rm f}}} \simeq {\cal 
N}+1 \, , 
\label{2e8} \\
\end{equation}
where
\begin{equation}
H_{{\rm f}} = \alpha^2 \sqrt{\frac{2\lambda_{{\rm b}} \kappa_0}{3}} \, . 
\label{2e9}
\end{equation}
So far the model has three free parameters, but only two of them are relevant 
for inflation, the brane tension $\lambda_{{\rm b}}$ and the self-interaction of 
the 
inflaton field $\alpha$. 

The amplitude of scalar density perturbations can be calculated for the 
exponential potential, and it is given by \cite{cope,sahni}
\begin{equation}
A^2_{\rm s} \simeq \frac{8}{75} \frac{V^4_{{\rm i}}}{\alpha^2 \lambda_{{\rm b}} 
m^4_{\rm 
Pl}} = \frac{128}{75} \alpha^6 ({\cal N}+1)^4 \frac{\lambda_{{\rm b}}}{m^4_{\rm 
Pl}} \, 
. \label{2e91}
\end{equation}
In the usual scenario of steep inflation, the values of the inflation parameters 
can be adjusted so that we get the observed value $A^2_{\rm s} \simeq 4 
\times 10^{-10}$. In particular, we can determine the value of the brane tension 
in terms of $\alpha$ as
\begin{equation}
\lambda_{{\rm b}} \simeq \frac{2.3 \times 10^{-10}}{\alpha^6} \frac{m^4_{\rm 
Pl}}{({\cal 
N}+1)^4} \, . \label{2e92}
\end{equation}
Notice that any deviation of $\lambda_{{\rm b}}$ from the value obtained in 
Eq.~(\ref{2e92}) would imply either a larger or smaller amplitude for the scalar 
spectrum.

As was pointed out in Ref.~\cite{sahni}, the number of $e$-foldings can be 
unambiguously determined for the exponential potential, for which we obtain 
${\cal N} \simeq 70$. This result is based on Eq.~(\ref{2e91}) and is 
independent of the value of $A^2_{{\rm s}}$.

\subsection{After inflation}
\label{two:two}

Soon after inflation ends, the brane term becomes unimportant and we 
recover the standard Friedman equation. Hence, the friction on the inflaton 
field diminishes and the inflaton energy is dominated by the kinetic term and 
the scalar energy density falls as stiff matter, $\rho_\phi = \rho^{(\rm 
kin)}_\phi (a_{\rm kin}/a)^6$. The epoch that follows is called the `kinetic 
epoch' or `kination' \cite{MJ2}, and we shall use `kin' to label the value of 
the different quantities 
at the beginning of the kinetic epoch. 

Since we are now working within standard cosmology (at energies $\rho < 
\lambda_{{\rm b}}$), the Hubble factor evolves as
\begin{eqnarray}
H = H_{\rm kin} \left(\frac{a_{\rm kin}}{a}\right)^3, \, && \, H^2_{\rm kin} 
\simeq \frac{\kappa_0}{3} \, 
\rho^{(\rm kin)}_\phi \, . \label{2e10}
\end{eqnarray}
The kinetic regime does not commence immediately after the end of inflation, but 
the values of the Hubble parameter at the end of inflation and at the beginning 
of the kinetic regime are related by the formula \cite{sahni}
\begin{equation}
\frac{H_{\rm kin}}{H_{{\rm f}}} = 0.085 -\frac{0.688}{\alpha^2} \, . 
\label{2e11}
\end{equation}

A small amount of radiation has been produced quantum-mechanically during 
inflation with an energy density $\rho_{\rm r} \sim 0.01 g_p H^4_{{\rm f}}$, 
where $g_p=10-100$ is the number of different species created from the vacuum. 
If the inflaton potential does not decay just after inflation, then this is the 
only mechanism to reheat the Universe. 

With the density of scalar matter falling more rapidly than that of radiation, 
the Universe 
continues evolving according to Eq.~(\ref{2e10}) until the small amount of 
radiation comes to dominate. Since the ratio of scalar matter to radiation at 
the end of inflation is large, $(\rho_\phi/\rho_{{\rm r}})_{{\rm f}} \gg 1$, 
radiation comes to dominate the expansion of the Universe only after a very long 
kinetic epoch. This is troublesome since the energy density of gravitational 
waves produced quantum-mechanically can exceed the nucleosynthesis constraints, 
as shown in Ref.~\cite{sahni}. 

In braneworld inflation, the amplitude of gravitational waves $h^2_{\rm GW}$ is 
enhanced with respect to the standard scenario, and its value at the end of 
inflation is given by
\begin{equation}
h^2_{\rm GW} \simeq 3 \alpha^2 ({\cal N}+1) \frac{H^2_{{\rm i}}}{m^2_{\rm Pl}} 
\, . 
\label{2e12}
\end{equation}
Gravitational waves behave as massless scalar fields and then their amplitude 
remains 
constant during inflation. During the kinetic epoch, the energy density of 
gravitational waves evolves as \cite{sahni}
\begin{equation}
\rho_g = \frac{32}{3\pi} h^2_{\rm GW} \rho_\phi \left( \frac{a}{a_{\rm kin}} 
\right)^2 \, . \label{2e13}
\end{equation}
From this, the energy density of gravitational waves at the time of scalar stiff 
matter--radiation equality \mbox{$(\rho_\phi=\rho_{{\rm r}})$} is
\begin{equation}
\left. \frac{\rho_g}{\rho_{{\rm r}}} \right|_{a = a_{\rm eq}} = \frac{64}{3\pi} 
h^2_{\rm 
GW} \left( \frac{a_{\rm eq}}{a_{\rm kin}} \right)^2 \simeq \frac{64}{3\pi} 
h^2_{\rm GW} \left. \frac{\rho_\phi}{\rho_{{\rm r}}} \right|_{{\rm f}} \gg 1 \, 
. 
\label{2e14}
\end{equation}
Thus, the energy density of gravitational waves on {\it short wavelength scales} 
\cite{sahni} exceeds the contribution of radiation. It happens then that 
gravitational waves come to dominate the stiff scalar matter well before 
radiation, which in fact never dominates, and then nucleosynthesis constraints 
are not satisfied at all. Therefore, the original steep inflation model seems to 
be 
ruled out \cite{sahni}.

However, the production of gravitational waves can be kept under control if the 
kinetic epoch is shortened. The latter can be done either by a sufficiently 
early 
decay of the inflaton field or by having a post-inflationary equation of state 
softer than that of stiff matter. The latter picture is quite appealing since 
the 
surviving inflaton field could become part of the dark matter at late 
times \cite{sahni,lidsey,FJ}.

\section{The curvaton field}
\label{three}

In this section we explore the possibility that the inflaton field is not 
responsible for providing the primordial fluctuations required for the formation 
of structure in the late Universe, as in the case where the value of the brane 
tension is lower than that in Eq.~(\ref{2e92}). Instead, the primordial 
fluctuations are generated by a second scalar field $\sigma$, usually called the 
{\it curvaton field}, through initially isocurvature perturbations \cite{lyth1}. 
For simplicity, we assume here that the curvaton field has a quadratic scalar 
potential $U(\sigma)=m^2 \sigma^2 /2$, and hence obeys the Klein--Gordon 
equation
\begin{equation}
\ddot{\sigma}+3H\dot{\sigma}+m^2\sigma = 0 \, , \label{3e1}
\end{equation}
where $m$ is the curvaton mass.

We will now follow the evolution of the curvaton field through different stages 
and find the constraints upon the free parameters of the model in order to have 
a viable curvaton scenario together with steep inflation. In doing this, we 
shall follow a similar procedure to that in Ref.~\cite{nicola}. We begin with a 
brief overview of the dynamics.

First of all, it is assumed that the curvaton field coexists with the inflaton 
field during inflation, during which the inflaton energy density is the dominant 
component. The curvaton field should survive the rapid expansion of the 
Universe, and for that it has to be effectively massless. As we shall see below, 
this condition is translated into a constraint of the curvaton mass $m$. If this 
constraint is satisfied, the curvaton field will remain constant to its initial 
value.

The following stage is that when the curvaton field becomes effectively massive, 
and this will generally happen during the kinetic epoch. In order to prevent a 
period of curvaton-driven inflation, the Universe must remain inflaton-dominated 
until this time. The latter condition imposes a constraint on the initial values 
the curvaton field can take, $\sigma_{{\rm i}}$. Once effectively massive, the 
curvaton field oscillates around the minimum of its potential and its energy 
density evolves as non-relativistic matter.

The final stage is that of curvaton decay into radiation, and the standard big 
bang cosmology is recovered afterwards. In the general case, curvaton decay 
should occur before nucleosynthesis, but other constraints arise depending on 
epoch of the decay, governed by the decay parameter $\Gamma_\sigma$. There are 
two scenarios to be considered, depending whether the curvaton field decays 
before or after it becomes the dominant component of the Universe. Whatever the 
case, the energy from gravitational waves should be maintained under control, 
which will actually become an extra constraint involving different parameters of 
the model. Also, we calculate the curvaton perturbations for each case which 
will help us to reduce the number of free parameters. At this point, the 
gaussianity conditions on the perturbations will be taken into account too. 

The introduction of the curvaton field adds three free parameters to the steep 
inflation model: the curvaton mass $m$, the initial value of the curvaton field 
$\sigma_{{\rm i}}$ and the curvaton decay parameter $\Gamma_\sigma$. However, a 
successful model needs to satisfy the conditions outlined above which are strong 
enough to restrict the allowed curvaton parameters. 

\subsection{Inflation}
\label{three:one}

We first study the evolution of the curvaton field during braneworld inflation 
with an exponential inflaton potential as described in Section~\ref{two}. We 
change the independent variable to $z \equiv 
3H/H_{{\rm f}}$, for which the equation of motion of the curvaton field 
Eq.~(\ref{3e1}) reads
\begin{equation}
z^2 \sigma^{\prime \prime} + z(1-z)\sigma^\prime + \frac{m^2}{H_{{\rm f}}^2} \, 
\sigma = 0 \, , 
\label{3e2}
\end{equation}
where primes denote derivatives with respect to $z$.

Assuming that the curvaton field is effectively massless during inflation $m \ll 
H_{{\rm f}}$, the last term in Eq.~(\ref{3e2}) is small compared to the others, 
so that the curvaton equation of motion is
\begin{equation}
z^2 \sigma^{\prime \prime} + z(1-z)\sigma^\prime \simeq 0 \, , \label{3e3}
\end{equation}
whose exact solutions are
\begin{eqnarray}
\sigma (z) &=& \sigma_{{\rm i}} + \sigma^\prime_{{\rm i}} z_{{\rm i}} 
e^{-z_{{\rm i}}}\left[ {\rm Ei}(z)-{\rm Ei}(z_{{\rm i}}) 
\right] \, , \label{3e4} \\
\sigma^\prime &=& \sigma^\prime_{{\rm i}}  \, \frac{z_{{\rm i}}}{z} \,  \exp 
\left(z-z_{{\rm i}} \right) \, . 
\label{3e5}
\end{eqnarray}
Here, $\sigma_{{\rm i}}$ and $\sigma^\prime_{{\rm i}}$ are the initial values of 
the 
field and its 
derivative, and ${\rm Ei}(z)$ is the exponential integral. In the case that 
${\cal 
N}=70$, so that $z$ runs in the range $z_{{\rm i}}=213 \rightarrow z_{{\rm 
f}}=3$ (see Eq.~(\ref{2e8})), the ratio of the final value  of the curvaton 
field to its 
initial value is
\begin{equation}
\frac{\sigma_{{\rm f}}}{\sigma_{{\rm i}}} = 1-1.005 \, \frac{\sigma^\prime_{{\rm 
i}}}{\sigma_{{\rm i}}} \, . 
\label{3e6}
\end{equation}

We see that the final value $\sigma_{{\rm f}}$ depends on the initial 
conditions. The interesting case is that in which the curvaton energy is equally 
distributed into kinetic and potential at the beginning of inflation, so that we 
can take $\dot{\sigma_{{\rm i}}}^2 \sim m^2 \sigma^2_{{\rm i}}$. Thus, from 
$\dot{\sigma_{{\rm i}}} = H_{{\rm f}} z_{{\rm i}} \sigma^\prime_{{\rm i}}$ we 
obtain
\begin{equation}
\frac{\sigma^\prime_{{\rm i}}(z)}{\sigma_{{\rm i}}(z)} \sim \frac{m}{H_{{\rm f}} 
z_{{\rm i}}} = \frac{1}{3({\cal N}+1)} \frac{m}{H_{{\rm f}}} \, . \label{3e7} 
\end{equation}
Therefore, the condition $m \ll H_{{\rm f}}$ makes the curvaton field remain 
constant 
under general initial conditions, and we can write $\sigma_{{\rm f}} \simeq 
\sigma_{{\rm i}}$. 
Notice also that $\dot{\sigma}_{{\rm f}} = 0$.

\subsection{Kinetic epoch}
\label{three:two}

As we anticipated at the beginning of this section, the dynamics of the curvaton 
field after inflation are different to the scenarios considered before in the 
literature. This difference is mainly because the curvaton field coexists with 
the surviving inflaton field, whose energy density is by far the dominant one at 
the end of inflation. The consequences of this will be explored in this section 
and our immediate concern will be to work out the corresponding gravitational 
wave constraint.

According to our hypothesis, it is during the kinetic epoch that the curvaton 
field becomes effectively massive. This happens at a time when $m \simeq H$, 
from which we obtain (see Eq.~(\ref{2e10}))
\begin{equation}
\frac{m}{H_{\rm kin}} = \frac{a^3_{\rm kin}}{a^3_{\rm m}} \, , 
\label{3e9}
\end{equation}
where `m' labels the quantities at the time when the curvaton becomes massive. 
Up to this point, the curvaton field remained effectively massless and we can 
consider that $\sigma_{{\rm m}} \simeq \sigma_{{\rm i}}$.

In order to prevent a period of curvaton-driven inflation, the Universe must 
still be dominated by the scalar stiff-matter at this point. Hence, using 
Eqs.~(\ref{2e10}) and (\ref{3e9}), we arrive to the restriction
\begin{equation}
\frac{m^2 \sigma^2_{{\rm i}}}{2 \rho^{\rm (m)}_\phi} = \frac{4\pi}{3} \frac{m^2 
\sigma^2_{{\rm i}}}{m^2_{\rm Pl} m^2} \ll 1 \, \Rightarrow \, \sigma^2_{{\rm i}} 
\ll 
\frac{3}{4\pi} m^2_{\rm Pl} \, . \label{3e10}
\end{equation}
Eq.~(\ref{3e10}) is sufficient to achieve subdominance at the time when the 
curvaton field becomes massive. But the curvaton energy should also be 
subdominant at the end of inflation, which implies
\begin{equation}
\frac{U_{{\rm f}}}{V_{{\rm f}}} = \frac{m^2 \sigma^2_{{\rm i}}}{4 \alpha^2 
\lambda_{{\rm b}}} \ll \frac{3 m^2 
m^2_{\rm Pl}}{16 \pi \alpha^2 \lambda_{{\rm b}}} = \alpha^2 \frac{m^2}{H^2_{{\rm 
f}}} \, 
, \label{3e11}
\end{equation}
where we have used Eqs.~(\ref{2e9}) and (\ref{3e10}). Thus, the curvaton mass 
should obey the stronger constraint 
\begin{equation}
m \ll \alpha^{-1} H_{{\rm f}} \, . \label{3e12}
\end{equation}
After the curvaton field becomes effectively massive, its energy decays as 
non-relativistic matter in the form
\begin{equation}
\rho_\sigma = \frac{m^2 \sigma^2_{{\rm i}}}{2} \frac{a^3_{{\rm m}}}{a^3} \, . 
\label{3e13}
\end{equation}
 
\subsection{Curvaton domination}
\label{three:three}

We calculate here the constraints to be applied on the curvaton 
model for the case in which the curvaton field comes to dominate the cosmic 
expansion, dealing first with the gravitational wave constraint. Since the 
curvaton energy redshifts as in Eq.~(\ref{3e13}), we find at the time of stiff 
scalar matter--curvaton matter equality $(\rho_\sigma = \rho_\phi)$ 
that\footnote{Note 
that the label $a_{\rm eq}$ has different meanings in 
Eqs.~(\ref{2e14}), (\ref{3e14}) and (\ref{3e20}).}
\begin{equation}
\left. \frac{\rho_\sigma}{\rho_\phi} \right|_{a=a_{\rm eq}} = \frac{4 \pi}{3} 
\frac{ m^2 \sigma^2_{{\rm i}}}{m^2_{\rm Pl} H^2_{\rm kin}} \frac{a^3_{{\rm 
m}}}{a^3_{\rm kin}} 
\frac{a^3_{\rm eq}}{a^3_{\rm kin}} = 1 \, , \label{3e14}
\end{equation}
Using Eqs.~(\ref{3e9}) and (\ref{3e14}) in Eq.~(\ref{2e13}), we can calculate 
the 
energy density of gravitational waves at this time, which should obey the 
constraint
\begin{equation}
\left. \frac{\rho_g}{\rho_\sigma} \right|_{a=a_{\rm eq}} = \frac{64}{3\pi} 
h^2_{\rm GW} \left( \frac{3}{4\pi} \frac{m^2_{\rm Pl}}{\sigma^2_{{\rm i}}} 
\frac{H_{\rm 
kin}}{m} \right)^{2/3} \ll 1 \, . \label{3e15}
\end{equation}

To finish with, we derive the conditions to be applied on the decay parameter 
$\Gamma_\sigma$. On the one hand, we require the curvaton field to decay before 
nucleosynthesis, and then $H_{\rm nucl}=10^{-40} m_{\rm Pl} < \Gamma_\sigma$. On 
the other hand, we also require the curvaton to decay after domination, and then 
$\Gamma_\sigma < H_{\rm eq}$, where $H_{\rm eq}$ is found by using 
Eqs.~(\ref{2e10}), (\ref{3e9}) and (\ref{3e14}):
\begin{equation}
H_{\rm eq} = H_{\rm kin} \frac{a^3_{\rm kin}}{a^3_{\rm eq}} = \frac{4 \pi}{3} 
\frac{\sigma^2_{{\rm i}}}{m^2_{\rm Pl}} m \, . \label{3e16}
\end{equation}
Hence, the constraint upon the decay parameter is
\begin{equation}
10^{-40} m_{\rm Pl} < \Gamma_\sigma < \frac{4 \pi}{3} \frac{\sigma^2_{{\rm 
i}}}{m^2_{\rm 
Pl}} m \, . \label{3e17}
\end{equation}

\subsection{Curvaton decay before domination}
\label{three:four}

We now turn our attention to the case in which the curvaton field decays 
before it dominates the cosmological expansion, but after it becomes massive so 
that we can use Eq.~(\ref{3e13}). As before, we calculate first the 
gravitational wave constraint. The curvaton field decays at a time when 
$\Gamma_\sigma = H$ and then
\begin{equation}
\frac{\Gamma_\sigma}{H_{\rm kin}} = \frac{a^3_{\rm kin}}{a^3_{{\rm d}}} \, , 
\label{3e18}
\end{equation}
where `d' labels the different quantities at the time of curvaton decay. 
The radiation produced from curvaton decay has an energy density 
\begin{equation}
\rho^{(\sigma)}_{r}= \frac{m^2 \sigma^2_{{\rm i}}}{2} \frac{a^3_{{\rm 
m}}}{a^3_{{\rm d}}} 
\frac{a^4_{{\rm d}}}{a^4} \, . \label{3e19} 
\end{equation}
We will now assume that this radiation energy density is much larger than that 
produced by inflation, a necessary requirement if we are to maintain the
gravitational 
waves under control. From Eq.~(\ref{3e19}), radiation equals the stiff
scalar 
matter $(\rho^{(\sigma)}_{{\rm r}} = \rho_\phi)$ at a time given by
\begin{equation}
\left. \frac{\rho^{(\sigma)}_{{\rm r}}}{\rho_\phi} \right|_{a=a_{\rm eq}} = 
\frac{4\pi}{3} \frac{m^2 \sigma^2_{{\rm i}}}{m^2_{\rm Pl} H^2_{\rm kin}} 
\frac{a^3_{{\rm m}}}{a^3_{\rm kin}} \frac{a^2_{\rm eq}}{a^2_{\rm kin}} 
\frac{a_{{\rm d}}}{a_{\rm 
kin}} = 1 \, . \label{3e20}
\end{equation}
This equation defines the ratio $a_{\rm kin}/a_{\rm eq}$ to be used in 
Eq.~(\ref{2e13}). Hence, the constraint from gravitational waves now reads
\begin{equation}
\left. \frac{\rho_g}{\rho_{{\rm r}}} \right|_{a = a_{\rm eq}} = \frac{16}{\pi^2} 
h^2_{\rm GW} \sqrt[3]{\frac{\Gamma_\sigma}{H_{\rm kin}}} \frac{m}{H_{\rm kin}} 
\frac{m^2_{\rm Pl}}{\sigma^2_{{\rm i}}} \ll 1 \, , \label{3e21}
\end{equation}
where we have also used Eqs.~(\ref{3e9}) and (\ref{3e18}).

Finally, we derive the new constraints on $\Gamma_\sigma$. The curvaton field 
should decay after it becomes massive, so that $\Gamma_\sigma < m$; and before 
it dominates the expansion of the Universe, $\Gamma_\sigma > H_{\rm eq}$ (see 
Eq.~(\ref{3e16})). Therefore, the constraints on the decay parameter are
\begin{equation}
\frac{4 \pi}{3} \frac{\sigma^2_{{\rm i}}}{m^2_{\rm Pl}} m < \Gamma_\sigma < m \, 
. 
\label{3e22}
\end{equation}

\subsection{The surviving inflaton field}

For completeness, we should add a small note here about the evolution of the 
inflaton field after curvaton domination and/or decay. It is well known that a 
steep exponential potential has a tracker behaviour where it evolves with the 
same equation of state as the dominant component \cite{scal}. However, during 
the kination regime, the field is fast-rolling, and this typically takes it to 
very low energy densities, so that it can join the tracker only late in the 
cosmological evolution. Detailed analysis, following Ref.~\cite{FJ}, would be 
needed to determine its precise evolution, but it is unlikely to have any 
significant cosmological consequences unless the exponential form of the 
potential becomes modified at low energies.

\subsection{Curvaton fluctuations}
\label{three:five}

Eqs.~(\ref{3e15}) and (\ref{3e22}) are the constraints on the free parameters of 
the 
model coming from gravitational waves, but their actual form is not useful yet 
because of the term $h^2_{\rm GW}$. The latter is indeed related to other 
parameters of the model through the curvaton fluctuations, as we shall show in 
this section.

The curvaton fluctuations $\delta \sigma_k$ obey the linearized Klein--Gordon 
equation
\begin{equation}
\ddot{\delta \sigma_k} + 3H \dot{\delta \sigma_k} + \left[ \frac{k^2}{a^2} + m^2 
\right] 
\delta \sigma_k = 0 \, , \label{3e23}
\end{equation}
where $k$ is the comoving wave number.

During the time the fluctuations are inside the horizon, they obey the same 
differential equation as the inflaton fluctuations do, for which we conclude 
that they acquire the amplitude $\delta \sigma_{{\rm i}} \simeq H_{{\rm 
i}}/2\pi$. Once the 
fluctuations are out of the horizon, they obey the same differential equation as 
the unperturbed curvaton field does and then we expect that they remain constant 
during inflation, under quite general initial conditions.

Up to this point, the evolution of the curvaton fluctuations resembles that of 
previous scenarios already present in the literature. But, in a similar manner 
to our analysis of the homogeneous curvaton field in Sections~\ref{three:three} 
and 
\ref{three:four}, we should now take into account that the inflaton field 
survives the inflationary period, and study how the final spectrum of 
perturbations 
could be modified by this. 

Following previous work, we can calculate the spectrum of the Bardeen parameter 
${\cal P}_\zeta$, whose observed value is about $2 \times 10^{-9}$. When decay 
occurs {\it after} curvaton domination, the produced perturbation 
is \cite{lyth1,lyth2,nicola}
\begin{equation}
{\cal P}_\zeta \simeq \frac{1}{9\pi^2} \frac{H^2_{{\rm i}}}{\sigma^2_{{\rm i}}} 
\, . 
\label{3e24}
\end{equation}

This case is by far the simplest one. The spectrum of fluctuations is 
automatically gaussian since $\sigma^2_{{\rm i}} \gg H^2_{{\rm i}}/4\pi^2$, and 
is independent of $\Gamma_\sigma$, a feature that will simplify the analysis of 
the parameter space. Moreover, even though the curvaton field coexists with the 
inflaton field up to this epoch, the spectrum of fluctuations is the same as in 
the standard scenario.

In the limit in which the curvaton decays when subdominant, the Bardeen 
parameter is given by \cite{lyth1,lyth2}
\begin{equation}
{\cal P}_\zeta \simeq \frac{r^2_{{\rm d}}}{36\pi^2} \frac{H^2_{{\rm 
i}}}{\sigma^2_{{\rm i}}} \, , 
\label{3e25}
\end{equation}
where the normalization takes into account that the dominant component at that 
time is the 
scalar stiff matter. Here $r_{{\rm d}}$ is the ratio of curvaton energy density 
to 
stiff 
scalar matter at curvaton decay, which is found by using Eqs.~(\ref{3e9}) and 
(\ref{3e18}) in
\begin{equation}
r_{{\rm d}} = \left. \frac{\rho_\sigma}{\rho_\phi} \right|_{a=a_{{\rm d}}} 
\!\!\! = 
\frac{4\pi}{3} 
\frac{m^2 \sigma^2_{{\rm i}}}{m^2_{\rm Pl} H^2_{\rm kin}} \frac{a^3_{{\rm 
m}}}{a^3_{\rm kin}} 
\frac{a^3_{{\rm d}}}{a^3_{\rm kin}} = \frac{4\pi}{3} \frac{m}{\Gamma_\sigma} 
\frac{\sigma^2_{{\rm i}}}{m^2_{\rm Pl}} \, . \label{3e26}
\end{equation}

Contrary to the previous case, Eqs.~(\ref{3e25}) and (\ref{3e26}) show 
explicitly the influence of the surviving inflaton field, and they are markedly 
different to the standard scenario in which the curvaton field coexists only 
with radiation.

\section{Model constraints}
\label{four}

In this section, we use the different constraints on the curvaton field to study 
under which conditions it can be used together with steep inflation. The two 
cases that were worked out before will be treated separately.

\subsection{Curvaton decay after domination}
\label{four:one}

In this case, Eqs.~(\ref{2e8}), (\ref{2e9}), and (\ref{3e24}) allow us to fix 
$\lambda_{{\rm b}}$ in terms of $\sigma_{{\rm i}}$ and $\alpha$ as
\begin{equation}
\frac{\lambda_{{\rm b}}}{m^4_{\rm Pl}} = \frac{27\pi}{16} \frac{{\cal 
P}_\zeta}{({\cal 
N}+1)^2 \alpha^4} \frac{\sigma^2_{{\rm i}}}{m^2_{\rm Pl}} \, , \label{4e1}
\end{equation}
and the amplitude of primordial gravitational waves Eq.~(\ref{2e12}) can be 
written in terms of ${\cal P}_\zeta$, $\sigma_{{\rm i}}$ and $\alpha$ as
\begin{equation}
h^2_{\rm GW} = 27 \pi^2 \alpha^2 ({\cal N}+1) {\cal P}_\zeta 
\frac{\sigma^2_{{\rm i}}}{m^2_{\rm Pl}} \, . \label{4e2}
\end{equation}
Thus, Eqs.~(\ref{3e15}) and (\ref{3e12}) are transformed into the following 
constraints on the curvaton mass:
\begin{eqnarray}
\frac{m}{m_{\rm Pl}} &\gg& 10^3 \sqrt{\pi^3 ({\cal N}+1)} \, \alpha \left( 2.6 
\alpha^2
 -21\right) {\cal P}^2_\zeta \frac{\sigma^2_{{\rm i}}}{m^2_{\rm Pl}} 
 \, , \;\;
\label{4e3} \\
\frac{m}{m_{\rm Pl}} &\ll& \frac{3\pi {\cal P}^{1/2}_\zeta}{\alpha ({\cal N}+1)} 
\frac{\sigma_{{\rm i}}}{m_{Pl}} \, , \label{4e4}
\end{eqnarray}
where we have used Eq.~(\ref{2e11}).

The initial value of the curvaton field $\sigma_{{\rm i}}$ is restricted by 
Eq.~(\ref{3e10}) and by another restriction coming from the assumption that the 
inflaton fluctuations are not relevant. The latter is obtained by using 
Eqs.~(\ref{2e92}) and (\ref{4e1}), which give 
\begin{equation}
\frac{\sigma_{{\rm i}}}{m_{\rm Pl}} < \frac{5}{\sqrt{72 \pi}} \frac{1}{\alpha 
({\cal N}+1)} \frac{A_{{\rm s}}}{\sqrt{{\cal P}_\zeta}} \, . \label{4e41}
\end{equation} 
In other words, Eq.~(\ref{4e41}) just means that the brane tension has a lower 
value than that of Eq.~(\ref{2e92}).

Finally, Eq.~(\ref{3e17}) restricts the value of the decay parameter 
$\Gamma_\sigma$, which can be transformed into another constraint upon 
$m$ and $\sigma_{{\rm i}}$ as
\begin{equation}
\frac{m}{m_{\rm Pl}} \frac{\sigma^2_{{\rm i}}}{m^2_{\rm Pl}} \gg \frac{3}{4\pi} 
\times 
10^{-40} \, . \label{4e5}
\end{equation}

\begin{figure}[t]
\includegraphics[width=8.5cm]{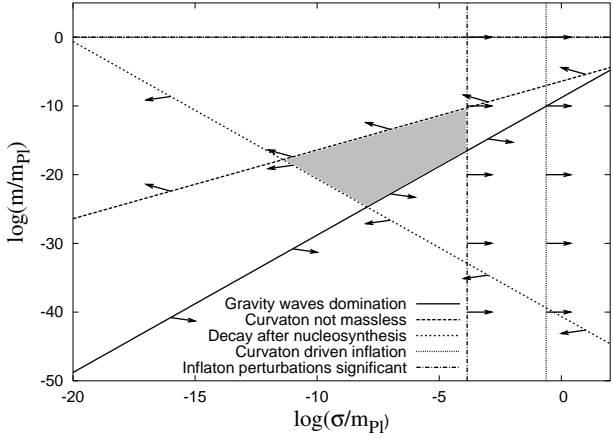}
\caption{\label{fig:f1} Allowed region of parameter space of the curvaton-steep 
inflation model for the case of curvaton domination before decay. Shown are the 
constraints Eqs.~(\ref{3e10}), (\ref{4e3}), (\ref{4e4}), (\ref{4e41}) and 
(\ref{4e5}) using $\alpha=15$ and ${\cal N}=70$. The regions excluded by each 
constraint are indicated by the arrows, and the allowed region is shaded. The 
allowed region is reduced for a larger value of $\alpha$.}
\end{figure}

The different conditions on the model of steep inflation $+$ curvaton field 
constrain the values of $\lambda_{{\rm b}}$, $\sigma_{{\rm i}}$ and $m$, but not 
of 
$\alpha$ and ${\cal N}$ which remain free parameters of the model. Therefore, we 
can 
draw 
the allowed region for the former in a plot $m$ versus $\sigma_{{\rm i}}$ by 
fixing the value of $\alpha$. Such a plot is shown in Fig.~\ref{fig:f1} for 
$\alpha=15$ and ${\cal N}=70$, in which we can see that there are viable models 
satisfying all the requirements.

However, the allowed region of the parameter space is reduced for larger values 
of the self-interaction of the inflaton field $\alpha \gg 1$, as we can see from 
Eqs.~(\ref{4e3}) and (\ref{4e4}). Thus, it is not always possible to find a 
successful model of curvaton field $+$ steep inflation, and this is mainly 
because of the gravitational wave constraint Eq.~(\ref{3e21}).

\subsection{Curvaton decay before domination}
\label{four:two}

For this case, it is convenient to use Eqs.~(\ref{3e25}) and (\ref{3e26}) to 
write $\sigma_{{\rm i}}$ in terms of the other quantities as 
\begin{equation}
\frac{\sigma^2_{{\rm i}}}{m^2_{\rm Pl}} = \frac{81}{4} \frac{{\cal 
P}_\zeta}{({\cal 
N}+1)^2} \frac{H^2_{\rm kin}}{H^2_{{\rm f}}} \frac{m^2_{\rm Pl}}{H^2_{\rm kin}} 
\frac{\Gamma^2_\sigma}{m^2} \, . \label{4e6}
\end{equation}
On the other hand, the amplitude of primordial gravitational waves 
Eq.~(\ref{2e12}) is rewritten as
\begin{equation}
h^2_{\rm GW} = 3 \alpha^2 ({\cal N}+1)^3 \frac{H^2_{{\rm f}}}{H^2_{\rm kin}} 
\frac{H^2_{\rm kin}}{m^2_{\rm Pl}} \, , \label{4e7}
\end{equation}
and then the gravitational wave constraint Eq.~(\ref{3e21}) reads
\begin{equation}
\frac{64 \alpha^2}{27\pi^2{\cal P}_\zeta} ({\cal N}+1)^5 \frac{H^4_{{\rm 
f}}}{H^4_{\rm 
kin}} \frac{H^4_{\rm kin}}{m^4_{\rm Pl}} \left( \frac{H_{\rm 
kin}}{\Gamma_\sigma} \right)^{5/3} \frac{m^3}{H^3_{\rm kin}} \ll 1 \, . 
\label{4e8}
\end{equation}
In the same manner, Eq.~(\ref{3e22}) is now written as
\begin{equation}
\frac{27 \pi {\cal P}_\zeta}{({\cal N}+1)^2} \frac{H^2_{\rm kin}}{H^2_{{\rm f}}} 
\frac{m^2_{\rm Pl}}{H^2_{\rm kin}} \frac{\Gamma^2_\sigma}{m} < \Gamma_\sigma < m 
\, . \label{4e9}
\end{equation}
Notice that, from Eqs.~(\ref{3e10}) and (\ref{4e9}) respectively, there are 
three 
similar constraints on the curvaton mass of the form
\begin{eqnarray}
\frac{m}{m_{\rm Pl}} &>& \left( \frac{\sqrt{27\pi {\cal P}_\zeta}}{{\cal N}+1} 
\frac{H_{\rm kin}}{H_{{\rm f}}} \frac{m_{\rm Pl}}{H_{\rm kin}} \right) 
\frac{\Gamma_\sigma}{m_{\rm Pl}} \, , \label{4e10} \\
\frac{m}{m_{\rm Pl}} &>& \left( \frac{\sqrt{27 \pi {\cal P}_\zeta}}{{\cal N}+1} 
\frac{H_{\rm kin}}{H_{{\rm f}}} \frac{m_{\rm Pl}}{H_{\rm kin}} \right)^2 
\frac{\Gamma_\sigma}{m_{\rm Pl}} \, , \label{4e11} \\
\frac{m}{m_{\rm Pl}} &>& \frac{\Gamma_\sigma}{m_{\rm Pl}} \, , \label{4e12}
\end{eqnarray}
while Eqs.~(\ref{3e12}) and (\ref{4e8}) give the following constraints, 
respectively,
\begin{eqnarray}
\frac{m}{m_{\rm Pl}} &\ll& \frac{1}{\alpha}\frac{H_f}{H_{\rm kin}} \frac{H_{\rm 
kin}}{m_{\rm Pl}} \, , \label{4e13} \\
\frac{m}{m_{\rm Pl}} &\ll& \left[ \frac{27\pi^2{\cal P}_\zeta}{64 \alpha^2 
({\cal N}+1)^5} \frac{H^4_{\rm kin}}{H^4_{{\rm f}}} \right]^{1/3} \left( 
\frac{m^8_{\rm 
Pl}}{H^8_{\rm kin}} \frac{\Gamma^5_\sigma}{m^5_{\rm Pl}} \right)^{1/9} \, . 
\label{4e14}
\end{eqnarray}

At this point, we should take into account the gaussianity condition upon the 
curvaton perturbations, since it is not automatically satisfied as in 
Section~\ref{four:one}. The gaussianity condition implies that the perturbations 
must be small compared to the mean value of the curvaton field, and then 
$\sigma^2_{{\rm i}} \gg H^2_{{\rm i}}/4\pi$. Using Eqs.~(\ref{2e8}), 
(\ref{2e11}) and (\ref{4e6}), we obtain
\begin{equation}
\frac{m}{m_{\rm Pl}} \ll \frac{9 \sqrt{\pi {\cal P}_\zeta}}{({\cal 
N}+1)^2} \frac{H^2_{\rm kin}}{H^2_{{\rm f}}} \frac{m^2_{\rm Pl}}{H^2_{\rm kin}}  
\frac{\Gamma_\sigma}{m_{\rm Pl}} \, . \label{4e15}
\end{equation}

All these constraints can be fulfilled if $m,\Gamma_\sigma \rightarrow 0$ 
independently of the value of $H_{\rm kin}$, but we know that this should not be 
possible since we still expect to recover the standard big bang scenario at 
nucleosynthesis. Hence, we will also impose the restriction $\Gamma_\sigma > 
10^{-40} m_{\rm Pl}$.

The final constraint has to do with the inflaton fluctuations, which should be 
negligible compared to the curvaton ones. In principle, this is a constraint 
upon $\lambda_b$ through Eq.~(\ref{2e92}), but Eqs.~(\ref{2e9}) and (\ref{2e11}) 
allow us to transform it into a constraint on $H_{\rm kin}$ in the form
\begin{equation}
\frac{H_{\rm kin}}{m_{\rm Pl}} \ll \frac{6.21 \times 10^{-5}}{\alpha^3 ({\cal 
N}+1)^2} \left( 0.085\alpha^2 -0.688 \right) \, . \label{4e17}
\end{equation}

\begin{figure}[t]
\includegraphics[width=8.5cm]{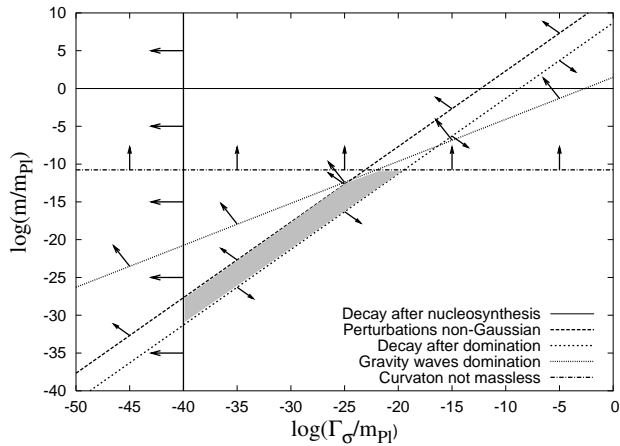}
\caption{\label{fig:f2} Curvaton constraints for the case in which the curvaton 
field decays before domination, using $\alpha=15$, ${\cal N}=70$ and $(H_{\rm 
kin}/m_{\rm Pl})=2.13 \times 10^{-11}$ in Eqs.~(\ref{4e8}), 
(\ref{4e10})--(\ref{4e15}), (\ref{4e17}) and the nucleosynthesis constraint. As 
in Fig.~\ref{fig:f1}, the regions excluded by each constraint are indicated by 
the arrows, and the allowed region is shaded. The allowed region would be 
smaller for a smaller (larger) value of 
$H_{\rm kin}$ ($\alpha$).}
\end{figure}

To unravel the parameter space allowed by the different constraints, we proceed 
as follows. We will take as free parameters the inflationary ones, $\alpha,{\cal 
N}$ and $H_{\rm kin}$, but the latter will take only values permitted by 
Eq.~(\ref{4e17}). Once we fix the values of $\alpha$, ${\cal N}$ and $H_{\rm 
kin}$ by 
hand, we plot only the strongest constraint of Eqs.~(\ref{4e10})--(\ref{4e12}) 
(the other two will be automatically satisfied) and the constraints in 
Eqs.~(\ref{4e13}), (\ref{4e14}) and (\ref{4e15}).
An example of this is shown in Fig.~\ref{fig:f2} for the values $\alpha=15$ and 
${\cal N}=70$, where we also assumed that inflaton fluctuations contribute with 
as much as the $10\%$ of the COBE signal. 

The different constraints permit some room for viable models, though the allowed 
region is smaller for smaller values of $H_{\rm kin}$, i.e.~smaller values of 
$\lambda_{{\rm b}}$, and for larger values of $\alpha$. Then, we conclude that a 
realistic scenario in which the curvaton field decays before domination is very 
restricted by the different constraints.

\section{Conclusions}
\label{five}

The steep inflation model is of interest as it ends inflation in a novel manner, 
through reduction of inflation-sustaining braneworld corrections, and because it 
generates inflation from the steep exponential potentials often found in 
supergravity phenomenology. Unfortunately, the original steep inflation model 
appears unable to match observations, due to an excessive amplitude of 
short-scale gravitational waves and because its perturbations are not close 
enough to scale-invariance.

In this paper we have shown that curvaton reheating can alleviate these 
difficulties, by dramatically raising the reheat temperature (as compared to 
gravitational particle production reheating), while simultaneously suppressing 
the large-angle gravitational wave contribution. The model is subject to a 
complex network of constraints, but viable models do exist, provided the brane 
tension $\lambda_{{\rm b}}$ and the exponential slope $\alpha$ are not too 
large. We should
stress that our analysis applies only to the case where the inflaton survives 
until late times, since an early decay of the inflaton field would lead to the 
standard scenario already studied in the literature \cite{lyth1,nicola,lyth2}.

An interesting feature of the steep inflation model is that the inflaton field 
may survive to the present, and with suitable modification of the potential from 
its exponential form at low energies it could act as either dark matter 
\cite{lidsey,FJ} or quintessence \cite{cope,PV,HL}. This resembles the scenarios
given in Ref.~\cite{lyth2} concerning residual isocurvature matter 
perturbations. For instance, if cold dark matter is created before curvaton 
decay, then there is a maximal correlation between curvature and 
CDM-isocurvature 
perturbations which is at variance with current observations.
We expect that the inflaton perturbations would be affected in the same way if 
the inflaton field is to be the dark matter at late times. The case of 
quintessence should not be troublesome, since the quintessence perturbations 
should decay upon horizon entry. However these situations merit more detailed 
investigation.


\begin{acknowledgments}
L.A.U.-L.~was supported by CONACyT, M\'exico under grant 010385, and A.R.L.~in 
part by the Leverhulme Trust.
\end{acknowledgments}


\end{document}